\documentclass[9pt, conference]{IEEEtran}
\IEEEoverridecommandlockouts
\usepackage{cite}
\usepackage{amsmath,amssymb,amsfonts}
\usepackage{algorithmic}
\usepackage{graphicx}
\usepackage{textcomp}
\usepackage{xcolor}
\usepackage{blindtext}
\usepackage{hyperref}
\usepackage{pgfplots}
\usepackage{xcolor}
\usepackage{booktabs} 
\usepackage{bbm}
\usepackage{bm}
\usepackage{multirow}
\usepackage{siunitx}
\usepackage{tikz}
\usetikzlibrary{shapes, arrows.meta, positioning}
\def\BibTeX{{\rm B\kern-.05em{\sc i\kern-.025em b}\kern-.08em
    T\kern-.1667em\lower.7ex\hbox{E}\kern-.125emX}}
\begin{document}

\title{MMMOS: Multi-domain Multi-axis Audio Quality Assessment}
\author{
Yi-Cheng Lin$^{\dagger}$, Jia-Hung Chen$^{\dagger}$, 
Hung-yi Lee
\\
 \textit{National Taiwan University}, \textit{$^\dagger$ Equal Contribution}\\

\texttt{\{f12942075, b10303106, hungyilee\}@ntu.edu.tw}\\
}
\maketitle

\begin{abstract}
Accurate audio quality estimation is essential for developing and evaluating audio generation, retrieval, and enhancement systems. Existing non-intrusive assessment models predict a single Mean Opinion Score (MOS) for speech, merging diverse perceptual factors and failing to generalize beyond speech. We propose MMMOS, a no-reference, multi-domain audio quality assessment system that estimates four orthogonal axes: Production Quality, Production Complexity, Content Enjoyment, and Content Usefulness across speech, music, and environmental sounds. MMMOS fuses frame-level embeddings from three pretrained encoders (WavLM, MuQ, and M2D) and evaluates three aggregation strategies with four loss functions. By ensembling the top eight models, MMMOS shows a 20–30\% reduction in mean squared error and a 4–5\% increase in Kendall’s $\tau$ versus baseline, gains first place in six of eight Production Complexity metrics, and ranks among the top three on 17 of 32 challenge metrics.
\end{abstract}

\section{Introduction}

Non‐intrusive speech quality assessment has been extensively studied, with models such as MOSA-Net \cite{mosanet}, Quality-Net \cite{quality_net}, and LDNet \cite{ldnet} proposed to predict a single MOS for speech signals. However, these approaches typically suffer from two key limitations: (1) they condense perceptual quality into a single scalar, obscuring important orthogonal aspects of quality; and (2) they are designed exclusively for speech, lacking the capacity to generalize to other audio domains such as music and environmental sounds.

This work presents our automatic quality assessment system \textbf{MMMOS}\footnote{\tiny\url{https://github.com/robert0518/MMMOS}} for AudioMOS Challenge 2025 track 2. This task aims to evaluate samples from text-to-speech (TTS), text-to-audio (TTA), and text-to-music (TTM) systems using 4 orthogonal axes to reduce the ambiguity of single-score evaluations. Production Quality (PQ) measures technical fidelity through clarity, dynamic range, frequency balance, and spatialization. Production Complexity (PC) quantifies the complexity of audio scenes by counting distinct sound components. Content Enjoyment (CE) captures the subjective aesthetic appeal in terms of emotional impact, artistic expression, and listener experience. Content Usefulness (CU) evaluates the suitability of samples as reusable material for content creation.

\begin{figure}[tbp]
  \centering
  \includegraphics[width=0.9\linewidth]{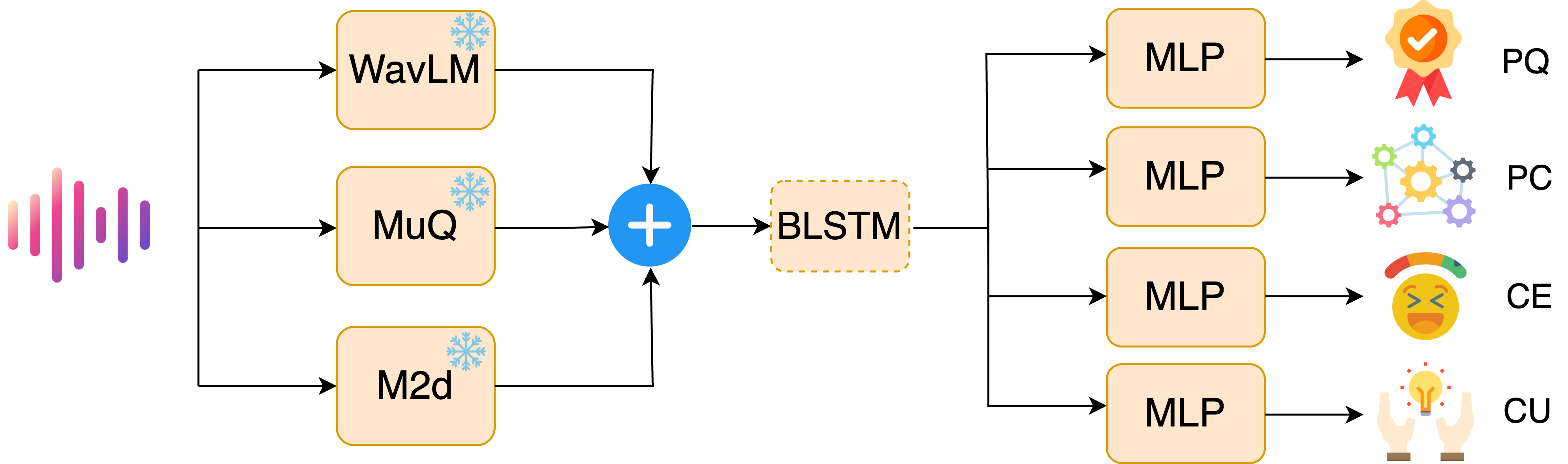}
  \caption{Model Architecture of \textbf{MMMOS}. BLSTM is an optional component depending on the aggregation method (Sec.~\ref{subsec:model}). }
  \label{fig:architecture}
  \vspace{-10pt}
\end{figure}

As depicted in Fig.~\ref{fig:architecture},  MMMOS leverages pretrained speech, audio, and music encoders for feature extraction. Then, the feature is processed by different feature aggregation modules and trained with different losses. Finally, the top-performing models on the development set are selected for ensembling. We achieved the 1st place for 6 out of 8 metrics in PC, and also the top 3 for 17 out of 32 evaluation metrics.
\vspace{-4pt}
\section{Related work}
Previous works often treat speech, music, and environmental sound as different domains for audio quality prediction. For speech audios, Quality-Net \cite{quality_net} and MOSNet \cite{mos_net} treated the Mean Opinion Scores (MOS) as a single regression target. Later, NISQA \cite{nisqa} and DNSMOS \cite{dnsmos} are learned to predict not only overall speech quality but also orthogonal metrics to give more interpretability. For music, previous works evaluate the audio quality of compressed music \cite{kasperuk2024non} or music heard through hearing aids \cite{haaqi}. 

For general audio, a GRU-based approach \cite{9664370} is first proposed to evaluate a single subjective quality score across speech, music, and environmental sounds. Audiobox Aesthetics trains a multi-faceted assessment model of audio clips across four dimensions using the WavLM \cite{wavlm} encoder, and calibrates the scores to be comparable for speech, music, and general audio. However, these works don't leverage the large-scale pretrained encoders from different domains. 

Some recent studies aim to learn a single audio representation that spans multiple domains. Wu et al. \cite{wu2022efficacy} proposed concatenating latent features from speech and pitch encoders to create a “holistic” embedding. Ritter-Gutierrez et al. \cite{distil_merge, permutation_merge} explore model-merging techniques to fuse speech and music encoders. While these methods improve general‐purpose audio representations, they havn't been applied to perceptual quality assessment.

\section{Methodology}

\subsection{Dataset}

We train and validate our models on two datasets: \textbf{AES-natural} \cite{audiobox_aesthetic}, the official challenge dataset, and \textbf{AES-PAM} \cite{pam}. AES-natural aggregates audio from real-world, public corpora covering speech (EARS \cite{ears}, LibriTTS \cite{libritts}, Common Voice 13.0 \cite{common}), music (MUSDB18‐HQ \cite{musdb18-hq}, MusicCaps \cite{musiccaps}), and sound effects (Audioset \cite{audioset}). It covers different audio types, qualities, and sampling rates. AES-PAM, in turn, consists of synthetic utterances generated by various TTA and TTS systems, mirroring the artificial nature of the challenge test set. To preserve each method’s representation in both sets, we stratified the data by generation model and then partitioned each stratum into training and development sets using an 80/20 split.

After downloading AES-natural’s clips, we exclude the unavailable ones from our splits, yielding 3,367 training and 434 development samples. All audio is resampled to 16 kHz mono. During training, any clip longer than 10 seconds is randomly cropped to a 10-second segment, providing varied temporal contexts.

\subsection{Model}
\label{subsec:model}
Each waveform is passed in parallel through three frozen encoders: WavLM Base+\footnote{\tiny \url{https://github.com/microsoft/unilm/tree/master/wavlm}}\cite{wavlm} for speech, MuQ\footnote{\tiny \url{https://huggingface.co/OpenMuQ/MuQ-large-msd-iter}}\cite{muq} for music, and M2D\footnote{\tiny \url{https://github.com/nttcslab/m2d/releases/download/v0.3.0/m2d_clap_vit_base-80x1001p16x16-240128_AS-FT_enconly.zip}}\cite{m2d} for general audio. We obtain utterance-level embeddings from each encoder by performing mean pooling over the temporal dimension of their hidden representations. The resulting embeddings are then normalized individually to ensure comparable scales across encoders, and concatenated along the feature axis to form the final fused representation.

We evaluate three aggregation strategies: \textbf{Direct aggregation (MLP)}: mean-pool the features across time, then map the embedding to per-axis scores via a lightweight MLP; \textbf{Hidden-state aggregation (BLSTM(h))}: mean-pool the features across time first, then feed it into a single-layer BLSTM and project the BLSTM’s final hidden state to per-axis scores; \textbf{Time-pooled aggregation (BLSTM(t))}: feed the concatenated frame-level sequence directly into BLSTM, mean-pool its outputs over time, and map to per-axis scores via MLP. The resulting embedding is passed to four independent MLPs to produce scalar MOS predictions. 

All models are trained with the Adam\cite{adam} optimizer at a learning rate of $10^{-4}$. Due to GPU VRAM limitations, BLSTM‐based aggregation models use a batch size of 16, whereas MLP models use a batch size of 32. We train for up to 10 epochs but stop early when the development‐set loss stops improving.

\subsection{Loss Function}

\subsubsection{Contrastive loss}

We use the contrastive loss (Con) \cite{cdpam} as a margin-based ranking objective to encourage the model to preserve the relative ordering of human-annotated scores. This loss penalizes predicted differences that exceed a specified margin, while ignoring minor deviations within the margin.

\subsubsection{UTMOS loss}

We use the UTMOS (UT) loss  \cite{utmos}, which combines a clipped mean squared error (MSE) regression loss with 0.5 times contrastive loss. The clipping parameter is set according to the original paper.

\subsubsection{Dual Criterion Quality loss} 
The Dual Criterion Quality (DCQ) Loss\cite{dcq} applies a pairwise objective over all sample pairs within a batch. It integrates a deviation term that penalizes mismatches between predicted and ground-truth score differences, alongside a ranking term that penalizes any inversion of the true ordering.

\subsubsection{Concordance Correlation Coefficient loss}
We use the Concordance Correlation Coefficient (CCC) loss \cite{cccloss} to penalize discrepancies in mean, variance, and covariance between predictions and ground truth, promoting both accurate and precise predictions.

\begin{table}[htbp]
\centering
\caption{Utterance‐ and system‐level performance for all teams on Production Complexity (best in \textbf{bold}, second in \underline{underline}).}
\setlength{\tabcolsep}{4pt}
\begin{tabular}{@{}l|cccc|cccc@{}}
\toprule
 & \multicolumn{4}{c|}{\textbf{Utterance-level}} & \multicolumn{4}{c}{\textbf{System-level}} \\
\textbf{Team} & \textbf{MSE} & \textbf{LCC} & \textbf{SRCC} & \textbf{KTAU} & \textbf{MSE} & \textbf{LCC} & \textbf{SRCC} & \textbf{KTAU} \\
\midrule
B02  & 0.562 & 0.905 & 0.902 & 0.723 & 0.226 & 0.963 & 0.934 & 0.800 \\
T03  & 0.876 & 0.920 & 0.907 & 0.730 & 0.573 & 0.981 & 0.944 & 0.816 \\
T04  & 0.553 & \underline{0.937} & \underline{0.921} & \underline{0.751} & 0.291 & \textbf{0.991} & 0.945 & 0.816 \\
T06  & 0.529 & 0.922 & 0.903 & 0.722 & 0.198 & 0.984 & 0.928 & 0.781 \\
T12  & 0.719 & 0.928 & 0.911 & 0.736 & 0.401 & 0.985 & 0.938 & 0.803 \\
T14  & 0.619 & 0.894 & 0.875 & 0.684 & 0.179 & 0.981 & 0.924 & 0.781 \\
T15  & 1.069 & 0.878 & 0.865 & 0.671 & 0.334 & \textbf{0.991} & \textbf{0.952} & \underline{0.822} \\
T20  & 0.561 & 0.889 & 0.870 & 0.679 & \textbf{0.112} & 0.981 & 0.918 & 0.756 \\
T21  & \underline{0.439} & 0.925 & 0.911 & 0.735 & \underline{0.135} & 0.987 & 0.942 & 0.806 \\
T24  & 0.535 & 0.888 & 0.877 & 0.689 & 0.203 & 0.947 & 0.897 & 0.743 \\
\midrule
Ours  & \textbf{0.432} & \textbf{0.942} & \textbf{0.922} & \textbf{0.752} & 0.168 & \textbf{0.991} & \underline{0.949} & \textbf{0.838} \\
\bottomrule
\end{tabular}
\label{tab:challenge}
\end{table}

\section{AudioMOS Challenge performance}
The challenge uses mean squared error (MSE) to quantify average prediction error, Pearson’s linear correlation coefficient (LCC) to measure linear agreement, Spearman’s rank correlation coefficient (SRCC) to assess monotonic consistency, and Kendall’s $\tau$ (KTAU) to capture ordinal alignment. These metrics are evaluated at the utterance level and system level. 

Our method demonstrates outstanding performance across both utterance- and system-level evaluations. Among the 10 participating systems, MMMOS gets the first place on 6 out of 8 metrics on PC (Table~\ref{tab:challenge}), and also wins the top-3 places on 17 out of 32 metrics. 

\begin{table}[htbp]
\centering
\caption{Utterance‐level performance of different encoder combinations.(Best in \textbf{bold}).}
\setlength{\tabcolsep}{4pt}
\begin{tabular}{@{}c c c c c c c@{}}
\toprule
\textbf{WavLM} & \textbf{MuQ} & \textbf{M2D} & \textbf{MSE} & \textbf{LCC} & \textbf{SRCC} & \textbf{KTAU} \\
\midrule
\checkmark & \checkmark & \checkmark & \textbf{0.338} & \textbf{0.840} & \textbf{0.847} & \textbf{0.676} \\
\checkmark &            & \checkmark & 0.339          & 0.831          & \textbf{0.847} & 0.673          \\
\checkmark & \checkmark &            & 0.480          & 0.766          & 0.776         & 0.597          \\
           & \checkmark & \checkmark & 0.371          & 0.829          & 0.843         & 0.671          \\
\checkmark &            &            & 0.997          & 0.367          & 0.391         & 0.278          \\
           & \checkmark &            & 0.799          & 0.520          & 0.509         & 0.356          \\
           &            & \checkmark & 0.355          & 0.826          & 0.842         & 0.669          \\
\bottomrule
\end{tabular}
\label{tab:encoder}
\end{table}

\begin{table*}[htbp!]
\centering
\small
\caption{Dev, PAM, and testing set performance for all feature/downstream/loss configurations. Agg.: Aggregation method. Ensemble: The models chosen for the submitted ensemble system. (Best in \textbf{bold}, second in \underline{underline}.)}
\setlength{\tabcolsep}{3 pt}
\begin{tabular}{@{}c c c c l l c
                 r r r r
                 r r r r
                 r r r r@{}}
\toprule
\# & \multicolumn{6}{c}{Configuration} 
  & \multicolumn{4}{c}{Dev Set (utterance level)} 
    & \multicolumn{4}{c}{PAM Set (utterance level)} 
      & \multicolumn{4}{c}{Test Set (system level)} \\
\cmidrule(lr){1-1} \cmidrule(lr){2-7} \cmidrule(lr){8-11} \cmidrule(lr){12-15} \cmidrule(lr){16-19}
 & WavLM & MuQ & M2D & Agg. & Loss & Ensemble
  & MSE & LCC & SRCC & KTAU 
  & MSE & LCC & SRCC & KTAU 
  & MSE & LCC & SRCC & KTAU \\
\midrule
1  & \checkmark &            & \checkmark & MLP    & CCC   &  
  & \underline{0.346} & \textbf{0.834} & \underline{0.846} & 0.662 & 
    0.339 & 0.900 & 0.906 & 0.731 & 
    1.366 & 0.918 & 0.879 & 0.737 \\
2  & \checkmark & \checkmark & \checkmark & MLP    & CCC   &  
  & 0.399 & 0.814 & 0.830 & 0.652 & 
    0.284 & 0.907 & 0.913 & 0.741 & 
    1.146 & 0.938 & 0.912 & 0.779 \\
3  & \checkmark & \checkmark & \checkmark & BLSTM(t)  & CCC   &  
  & 0.351 & 0.833 & 0.842 & 0.660 & 
    0.277 & 0.906 & 0.906 & 0.732 & 
    1.307 & 0.937 & 0.901 & 0.760 \\
4  & \checkmark & \checkmark & \checkmark & BLSTM(h)  & CCC   &  
  & 0.406 & 0.810 & 0.815 & 0.638 & 
    0.301 & 0.903 & 0.909 & 0.739 & 
    1.360 & 0.919 & 0.888 & 0.746 \\
5  & \checkmark &            & \checkmark & MLP    & Con   & \checkmark 
  & 0.390 & 0.825 & 0.839 & 0.663 & 
    0.329 & 0.912 & 0.915 & 0.746 & 
    1.679 & 0.930 & 0.890 & 0.733 \\
6  & \checkmark & \checkmark & \checkmark & MLP    & Con   &  
  & 0.415 & 0.813 & 0.820 & 0.645 & 
    0.329 & 0.907 & 0.912 & 0.739 & 
    1.576 & 0.939 & 0.916 & 0.789 \\
7  & \checkmark & \checkmark & \checkmark & BLSTM(t)  & Con   &  
  & 0.389 & 0.817 & 0.828 & 0.651 & 
    \textbf{0.223} & 0.914 & 0.914 & 0.742 & 
    \textbf{0.670} & 0.938 & 0.909 & 0.774 \\
8  & \checkmark & \checkmark & \checkmark & BLSTM(h)  & Con   & \checkmark 
  & 0.347 & 0.827 & 0.838 & \underline{0.668} & 
    \underline{0.229} & 0.917 & 0.916 & 0.750 & 
    0.986 & 0.939 & 0.912 & 0.777 \\
9  & \checkmark &            & \checkmark & MLP    & DCQ   & \checkmark 
  & 0.421 & 0.817 & 0.832 & 0.650 & 
    0.329 & 0.907 & 0.911 & 0.741 & 
    1.617 & 0.921 & 0.887 & 0.744 \\
10 & \checkmark & \checkmark & \checkmark & MLP    & DCQ   &  
  & 0.448 & 0.818 & 0.824 & 0.645 & 
    0.279 & \textbf{0.918} & \textbf{0.921} & \textbf{0.754} & 
    1.466 & 0.944 & \underline{0.919} & \textbf{0.796} \\
11 & \checkmark & \checkmark & \checkmark & BLSTM(t)  & DCQ   & \checkmark 
  & 0.564 & 0.824 & 0.832 & 0.656 & 
    0.405 & \underline{0.917} & 0.916 & 0.748 & 
    \underline{0.716} & \textbf{0.947} & 0.915 & 0.779 \\
12 & \checkmark & \checkmark & \checkmark & BLSTM(h)  & DCQ   &  
  & 0.515 & 0.822 & 0.831 & 0.650 & 
    0.427 & 0.903 & 0.900 & 0.724 & 
    0.974 & 0.939 & 0.909 & 0.773 \\
13 & \checkmark &            & \checkmark & MLP    & UT & \checkmark 
  & 0.361 & 0.822 & \textbf{0.849} & \textbf{0.672} & 
    0.298 & 0.910 & 0.911 & 0.745 & 
    1.163 & 0.919 & 0.872 & 0.726 \\
14 & \checkmark & \checkmark & \checkmark & MLP    & UT & \checkmark 
  & \textbf{0.342} & \underline{0.833} & 0.838 & 0.656 & 
    0.268 & 0.909 & 0.912 & 0.744 & 
    1.112 & 0.940 & 0.913 & 0.782 \\
15 & \checkmark & \checkmark & \checkmark & BLSTM(t)  & UT & \checkmark 
  & 0.378 & 0.827 & 0.832 & 0.656 & 
    0.327 & 0.912 & 0.914 & 0.747 & 
    1.397 & 0.925 & 0.893 & 0.746 \\
16 & \checkmark & \checkmark & \checkmark & BLSTM(h)  & UT & \checkmark 
  & 0.358 & 0.823 & 0.836 & 0.653 & 
    0.245 & 0.916 & \underline{0.919} & \underline{0.752} & 
    1.113 & \underline{0.945} & \textbf{0.920} & \underline{0.793} \\
\bottomrule
\end{tabular}
\label{tab:aggregated}
\vspace{-6pt}
\end{table*}

\section{Result}

\subsection{Ablation on Encoders}
To isolate the effect of encoder fusion, all combinations of encoders were evaluated on the dev set using the UTMOS loss and MLP aggregation to produce the MOS predictions. For each utterance, we average the four axis-specific outputs (CE, CU, PC, PQ) into a single composite score, then compute four evaluation metrics.

From Table~\ref{tab:encoder}, both WavLM+M2D and WavLM+M2D+MuQ attain the top SRCC of 0.847. Adding MuQ to WavLM+M2D further reduces MSE and increases LCC and KTAU, indicating more accurate and consistent predictions without sacrificing rank correlation. By contrast, MuQ+M2D and M2D alone deliver slightly lower correlation and higher error, while WavLM and MuQ individually perform poorly.


Because our primary objective is to maximize utterance‐level SRCC while maintaining low MSE and strong LCC and KTAU, we select the WavLM+M2D (BiEnc) and WavLM+M2D+MuQ (TriEnc)  configurations for further exploration. 


\subsection{Ablation on Adaptor}
From Table \ref{tab:encoder},  we choose the BiEnc and TriEnc as our base encoders. Moreover, the TriEnc yielded the best overall metrics; accordingly, we selected this configuration for further experiments. Fixing these encoders, we then evaluated four downstream aggregation schemes: (1) MLP on BiEnc features, (2) MLP on  TriEnc features, (3) BLSTM(h) on TriEnc features, (4) BLSTM(t) on TriEnc features. The models are trained with the UTMOS loss.

We use utterance‐level SRCC on the Dev set as our model‐selection criterion. As shown in Table~\ref{tab:aggregated}, the four downstream aggregation schemes (\#13-16) differ by less than 0.02 in Dev set SRCC, and vary by under 0.01 in PAM dev set SRCC. Given these minimal differences, we include all four methods in subsequent experiments. On the unseen test set, the hidden‐state aggregation exhibits the strongest generalization, while the linear aggregator applied to BiEnc alone achieves only 0.872 system‐level SRCC, indicating comparatively poorer performance.

\subsection{Ablation on Loss}
Examining Table~\ref{tab:aggregated}, we observe that with the BiEnc backbone, the UTMOS loss achieves the highest utterance‐level SRCC on the Dev set (0.849). Under the TriEnc backbone, UTMOS loss consistently ranks first or second across all aggregation methods.However, on the unseen test set, UTMOS yields the top system‐level SRCC in only one aggregation configuration and falls behind DCQ and Con in the others, suggesting that its generalization performance is poor. In contrast, the DCQ and Con losses, although modest on the Dev set, generalize more robustly, attaining equal or better system‐level SRCC on the test set across all aggregators. 

We also find that appending MuQ to the encoder invariably boosts test‐set SRCC compared to configurations without MuQ. This improvement confirms that music-domain embeddings provide complementary information that strengthens the model’s ability to generalize across both natural and synthetic audio samples.  

Finally, the ordering of aggregation schemes by SRCC on the PAM dev set is highly predictive of their test performance: the two highest‐ranked methods on PAM remain the top two on the unseen test set. This strong correlation implies that PAM dev set results serve as a reliable proxy for generalization, since both PAM dev set and the test set consist entirely of synthetic audio samples, while the Dev set comprises real recordings.  

\subsection{Ensemble Strategy}
\begin{table}[htbp!]
\centering
\caption{Comparison of ensemble strategies on the PAM dev and test sets.(best in \textbf{bold}.)}
\label{tab:ensemble}
\setlength{\tabcolsep}{3.5pt}
\begin{tabular}{@{}lcccccccc@{}}
\toprule
\textbf{Ensemble} & \multicolumn{4}{c}{\textbf{PAM Set}} & \multicolumn{4}{c}{\textbf{Test Set}} \\
\cmidrule(lr){2-5} \cmidrule(lr){6-9}
                  & MSE    & LCC    & SRCC   & KTAU   & MSE    & LCC    & SRCC   & KTAU   \\
\midrule
Submitted      & 0.228	& 0.921& 	0.923&	0.760  & 1.137  & 0.937  & 0.910  & 0.768  \\
All models             & 0.233&	0.920&	0.920&	0.755  & 1.140  & 0.937  & 0.914  & 0.781  \\
PAM top 8            & 0.212& 0.923  &	0.924&	0.760  & 1.046	& 0.942	& 0.917	& 0.784  \\
PAM top 4            & \textbf{0.204} &\textbf{0.925}	& \textbf{0.925}	& \textbf{0.761}  & \textbf{0.964}	& \textbf{0.946}& 	\textbf{0.923}& 	0.791  \\
Dev top 8 &0.212&	0.923&	0.924&	0.760&	1.251&	0.934&	0.906&	0.766\\
PAM top 1       & 0.279 &0.918& 0.921& 0.754 &1.466 &0.944 &0.919& \textbf{0.796}  \\
\bottomrule
\end{tabular}
\end{table}

Ensembling utilizes the complementary strengths of individual models to mitigate prediction variance and enhance robustness, thereby producing more stable and reliable quality estimates compared to any single configuration. In this work, we construct our submitted system by ensembling the intersection of the top 12 models ranked on the Dev and the top 12 models ranked on the PAM dev set, resulting in 8 models. For ablation, we compare four strategies: averaging all sixteen models, averaging the top-8 or top-4 SRCC models from the PAM dev set, and averaging the top-8 SRCC models from the development set. We use the best single model on the PAM dev set as our baseline.

The ablation results show that including PAM set performance in the model selection criterion produces a more robust ensemble than relying only on the development set. Selecting models solely by development set performance can inadvertently include those that overfit to natural recordings, leading to the worst ranking performances on the test set. By combining complementary configurations, \textbf{PAM top 4} consistently boosts correlation metrics on both the PAM set and on an unseen test set. At the same time, incorporating too many models can harm performance by introducing weaker predictors whose higher variance reduces overall accuracy.


\section{Conclusion}
We present a unified, non-intrusive audio quality assessment system that fuses features from three pretrained encoders, evaluates three aggregation strategies and four loss functions, and ensembles top-performing models. In the AudioMOS 2025 Track 2 challenge, our system placed first on 6 of 8 production-complexity metrics and placed top-3 on 17 out of 32 metrics. Compared to the official baseline, we achieve a 20–30 \% reduction in MSE and a 4–5 \% increase in KTAU across all four perceptual axes.

We prove that the integration of complementary encoders derived from speech, music, and general audio domains substantially improves the robustness and generalizability of quality prediction models. Also, the selective ensembling of models with high SRCC offers an effective trade-off between predictive performance and computational efficiency during inference. Collectively, these findings provide a comprehensive framework that can guide future development of more accurate audio quality prediction systems.

\bibliographystyle{IEEEtran}
\bibliography{mybib}

\end{document}